\documentclass[aps,reprint,graphicx,superscriptaddress]{revtex4-1}
\usepackage{graphicx}
\usepackage{subfigure}
\usepackage{hyperref}
\usepackage{amsmath}
\usepackage{amssymb}
\usepackage{epstopdf}
\usepackage{bbm}
\usepackage{threeparttable}
\usepackage{appendix}

\begin{document}

\title{Highly sensitive detection for infrared photons by non-degenerate two-photon absorption under mid-infrared pumping}

\author{Jianan Fang}
\affiliation{State Key Laboratory of Precision Spectroscopy, East China Normal University, Shanghai 200062, China}

\author{Yinqi Wang}
\affiliation{State Key Laboratory of Precision Spectroscopy, East China Normal University, Shanghai 200062, China}

\author{Ming Yan}
\affiliation{State Key Laboratory of Precision Spectroscopy, East China Normal University, Shanghai 200062, China}

\author{E Wu}
\affiliation{State Key Laboratory of Precision Spectroscopy, East China Normal University, Shanghai 200062, China}

\author{Kun Huang}
\email{khuang@lps.ecnu.edu.cn}
\affiliation{State Key Laboratory of Precision Spectroscopy, East China Normal University, Shanghai 200062, China}

\author{Heping Zeng}
\email{hpzeng@phy.ecnu.edu.cn}
\affiliation{State Key Laboratory of Precision Spectroscopy, East China Normal University, Shanghai 200062, China}
\affiliation{Jinan Institute of Quantum Technology, Jinan, Shandong 250101, China}
\affiliation{CAS Center for Excellence in Ultra-intense Laser Science, Shanghai 201800, China}
\affiliation{Shanghai Research Center for Quantum Sciences, Shanghai 201315, China}

\date{\today}

\begin{abstract}
We have demonstrated highly-sensitive photon counting in the infrared based on the two-photon absorption (2PA) in a silicon avalanche photodiode, where the required photon energy for inducing effective conductivity was provided by an intense mid-infrared (MIR) field at 3 $\mu$m. The used MIR pumping scheme could not only benefit from the enhanced 2PA coefficient in the non-degenerate regime, but also eliminate the detrimental background noises due to the pump harmonic excitation of the pump. Consequently, the enhancement factor for the signal counting rate unprecedented reached to about $10^{5}$ with input infrared pulses at the femtojoule level. Additionally, the noise equivalent power was substantially improved by two orders of magnitude comparing to conventional schemes with near-infrared pumping. Therefore, the presented configuration might provide an alternative to realize sensitive infrared detection and imaging with desirable features of room-temperature operation, no phase-matching requirement, and broadband responding window, which would find a variety of applications including remote ranging, sensitive sensing, biochemical imaging, and trace spectroscopy.
\end{abstract}

\maketitle

\section{Introduction}
Sensitive infrared detection plays an enabling role in numerous applications such as laser ranging, optical communication, environmental monitoring, biomedical sensing, and photoluminescence analysis \cite{Eisaman2011RSI,Zhang2015LSA}. Generally, the sensitivity of infrared detectors is limited by the relatively large dark noise especially at the room-temperature operation, which is partially due to the narrow semiconductor band gap \cite{Rogalski2017RPP}. This issue would become more conspicuous for optical detectors working at longer infrared wavelengths. Cryogenic operation could help to suppress the thermal radiation and dark current, yet it remains challenging to approach single-photon sensitivity \cite{Adiyan2019NC, Korzh2020NP}. In comparison, high-efficiency and low-noise photon counters are commercially available with silicon-based avalanche photodiodes (Si-APDs) due to their higher band gap \cite{Hadfield2009NP}. In this context, the so-called frequency upconversion technique has recently attracted increasing attention, where the infrared signal is spectrally translated into the visible band before being detected by a high-performance Si detector \cite{Huang1992PRL, Tanzilli2005Nature, Kang2020PTL}. 

This simple but effective protocol has mainly been instantiated in two categories. The first one relies on the nonlinear manipulation on the infrared field, such as resorting to sum frequency generation \cite{Pelc2011OE, Huang2012APL, Barh2019AOP} and four-wave mixing \cite{McGuinness2010PRL, Li2016NP}. In this case, phase-matching condition is typically required to realize efficient parametric interactions, yet usually at the price of narrow accepted wavelength window \cite{Mrejen2020LPR}. In parallel, the other strategy is based on the material nonlinearity of the detector itself, which specifically involves the manipulation of free charge carriers by two-photon \cite{Bristow2007APL, Reichert2016PRL} or multi-photon \cite{Hurlbut2007OL, Pearl2008APL, NevetOL2011} absorption. The underlying generation and recombination of charge carriers do not require phase matching, thus favoring broadband operation and free of optics alignment \cite{Hayat2008PRB}. The degenerate two-photon absorption (D-2PA) has long been recognized useful to detect infrared photons in wide-bandgap semiconductors. For instance, auto-correlators based on 2PA were commonly used to characterize ultrafast infrared pulses in the infrared regions \cite{Boiko2017APL}. Recently, extreme non-degenerate 2PA (ND-2PA) has been investigated to show a significant enhancement for the absorption coefficient comparing to the D-2PA case \cite{Fishman2011NP, Cirloganu2011OE, Poulvellarie2018PRAp, Zhang2015OE}. Consequently, sensitive mid-infrared imaging was demonstrated by using a direct-bandgap GaN photodiode sensor \cite{Pattanaik2016OE} and indirect-bandgap Si CCD camera \cite{Knez2020LSA}. Additionally, infrared photon counting has been demonstrated based on a GaAs photomultiplier tube (MPT) \cite{Boitier2009APL} and a Si-APD \cite{Xu2019PTL}. 

However, in the reported demonstrations, the detector sensitivity was inevitably limited by the D-2AP of the intensive pumping field. Although the sensitivity could typically be improved by using the lock-in detection with a modulated signal, the severe pump-induced noise may saturate the detector, thus limiting the dynamic range of the detector \cite{Fishman2011NP, Knez2020LSA}. To go beyond the demonstrated sensitivity, lower pump frequency could be chosen to eliminate the severe background noises due to the harmonic excitation of pump photons, whereas the parasitic process of three-photon absorption (3PA) is many orders of magnitude weaker and may be neglected \cite{Hayat2008PRB, Wang2013OE, Benis2020Optica}.

\begin{figure*}[t!]
\centering
\includegraphics[width=0.9\textwidth]{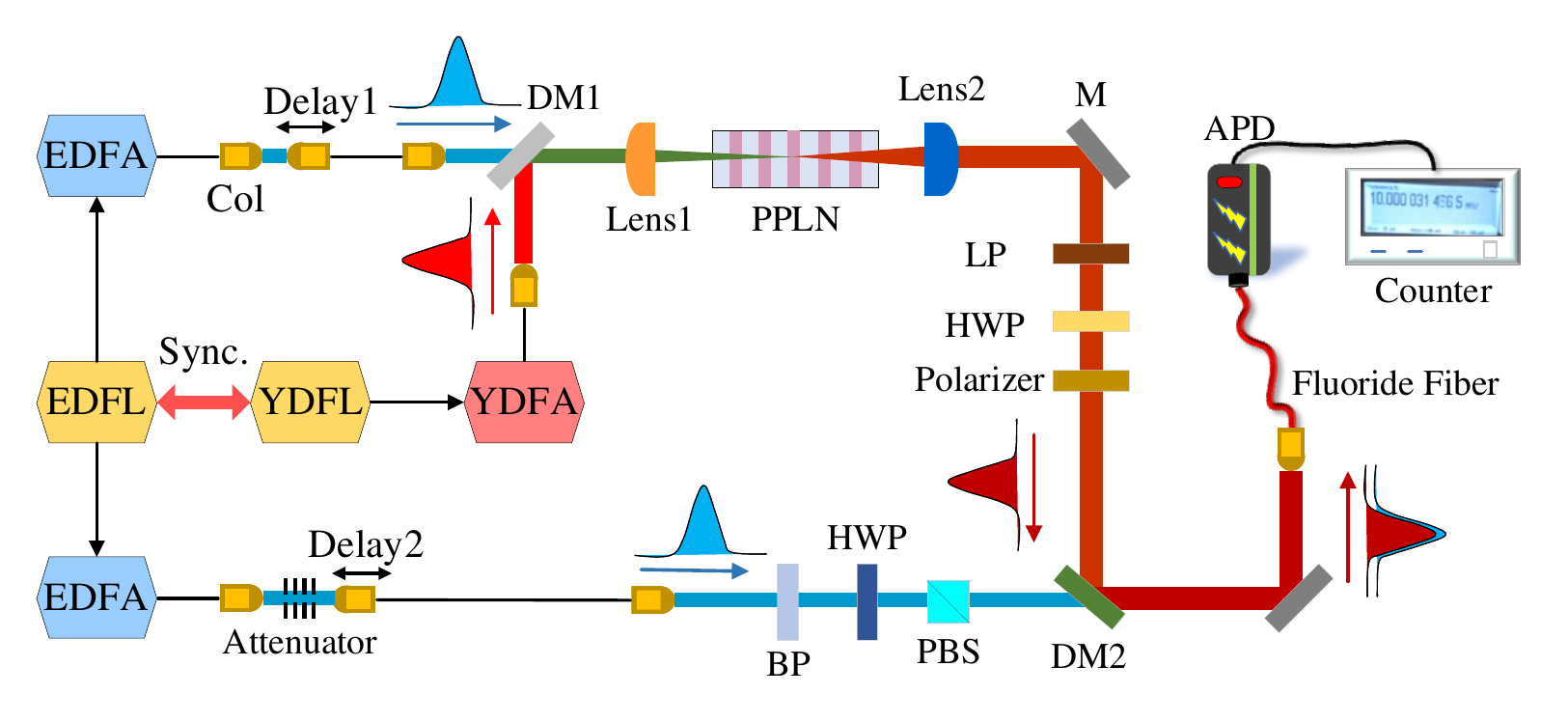}
\caption{Schematic of 2PA-based infrared detection based on temporally synchronized fiber lasers. The synchronized pulses at 1030 nm and 1550 nm originate from two Yb- and Er-doped fiber lasers (YDFL and EDFL), respectively. The pump source at MIR is prepared by the difference-frequency generation between the dual-color beams, while the signal source is obtained after a series of calibrated attenuators. The signal and pump sources are then spatially coupled into a section of fluoride fiber before being detected by a Si-APD. Col: collimator; BP: band-pass filter; LP: long-pass filter; HWP: half-wave plate; PBS: polarization beam splitter; DM: dichroic mirror; PPLN: periodically-poled lithium niobate crystal; APD: avalanche photodiode.}
\label{fig1}
\end{figure*}

Here, we demonstrated highly sensitive detection for infrared photons by ND-2PA in a Si-APD under mid-infrared (MIR) pumping. The chosen pump wavelength at 3070 nm made the pump photon energy of 0.4 eV much lower than the semiconductor mid-gap about 0.56 eV, which thus eliminated the D-2PA in spite of the high pump intensity. As a result, the residual background noise was mainly ascribed to the negligible three-photon absorption process, which enabled to improve the noise equivalent power by two orders of magnitude than that in conventional schemes with near-infrared pumping. Thanks to the substantial suppression of background noise, infrared pulses with femtojoule energy could be identified, which exhibited an enhanced counting rate by a factor closed to $10^{5}$ comparing to the case of direct detection based on D-2PA. Additionally, we experimentally investigated the thermal effect on the 2PA behavior, which showed a degradation of count rates as increasing the MIR pump power. This finding may provide useful guidance to employ the presented configuration of infrared detection in subsequent applications.

\section{Basic principles}
The underlying mechanism of the infrared detection scheme is based on absorbing two photons in a semiconductor with a band gap of $E_g$. The photon energy of signal and pump fields is $\hbar \omega_1$ and $\hbar \omega_2$. $\hbar \omega_1 + \hbar \omega_2 >  E_g$ should be satisfied to generate electro-hole pairs, thus giving rise to a photocurrent. In the non-degenerate scenario, the carrier density has a linear dependence on the pump power. On one hand, efficiency detection requires sufficiently high pump power. On the other hand, intensive pump intensity inevitably induces background noises due to harmonic absorption processes. The trade-off could be mitigated by choosing a lower pump photon energy. For instance, the condition of $\hbar \omega_2 < E_g / 2$ would exclude the appearance of D-2PA for the pump. Consequently, the pump-induced noise is substantially suppressed due to the much weaker 3PA process, hence leading to an enhanced sensitivity of the 2PA-based detector. Furthermore, theoretical investigation has revealed that infrared single-photon detection is possible in a Si-APD if the pump frequency is small enough to avoid the unwanted higher-order absorption processes \cite{Hayat2008PRB}.

\begin{figure*}[t!]
\centering
\includegraphics[width=0.8 \textwidth]{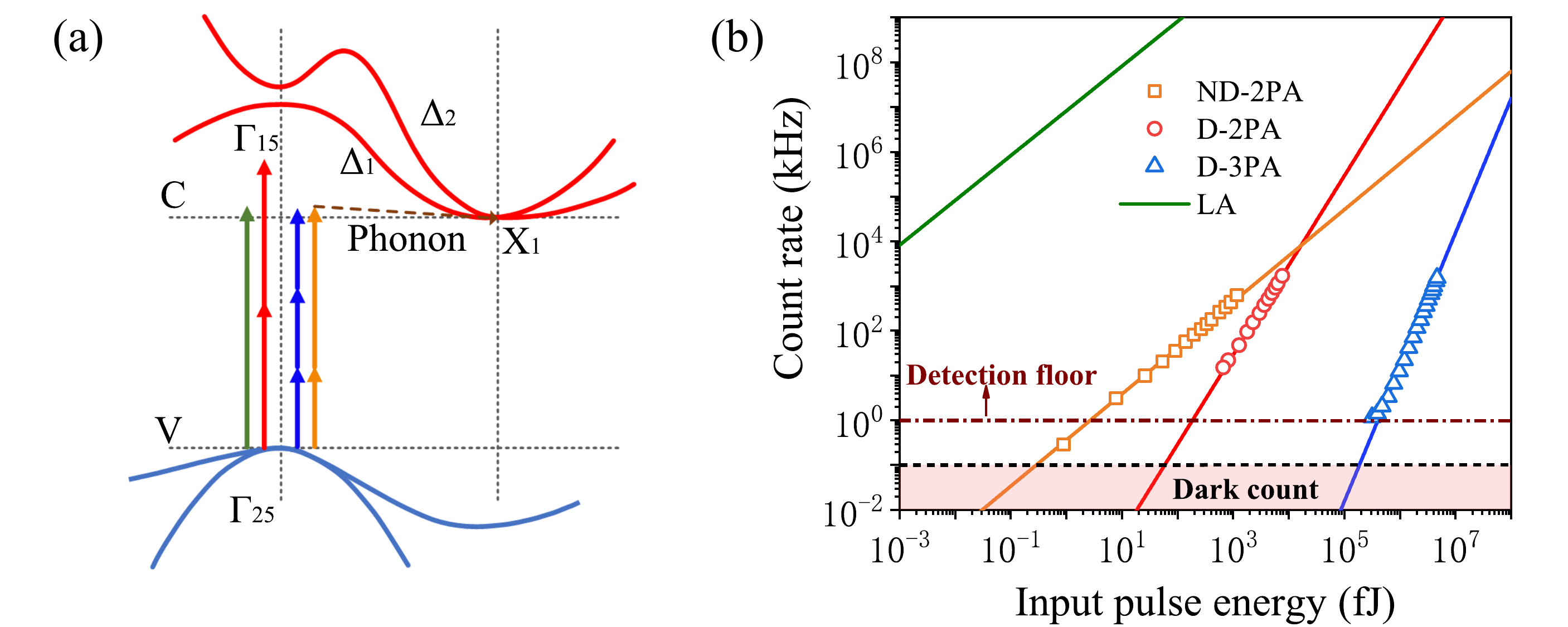}
\caption{Highly-sensitive infrared photon counting via two-photon absorption under mid-infrared pumping. (a) Energy diagram of an indirect bandgap silicon detector. The relevant transitions for comparison include linear absorption (LA) at 1030 nm, degenerate two-photon absorption (D-2PA) at 1550 nm, degenerate three-photon absorption (D-3PA) at 3070 nm, and non-degenerate two-photon absorption (ND-2PA). In the presence of intense mid-infrared pump, the infrared photon could promote a valence-band electron to a conduction-band state with the help of an intermediate photon. (b) Recorded count rates by the Si-APD as a function of input pulse energy for various energy-level transitions. Note that the MIR pulse energy at 3070 nm was set at 0.32 nJ for the non-degenerate curve. The dark count of 100 Hz is measured by blocking the injecting light onto the detector. The detection floor about 1 kHz is defined by the noise level with the presence of 0.32-nJ MIR pump pulses.}
\label{fig2}
\end{figure*}

Notably, the non-degenerate operation could significantly enhance the nonlinear absorption \cite{Fishman2011NP, Cirloganu2011OE}. Specifically, in the scattering matrix formalism with two parabolic bands, the 2PA coefficient $\alpha_2$ for direct bandgap semiconductors can be written as \cite{SheikBahae1991JQE}: 
\begin{equation}
\alpha_2(\omega_1, \omega_2) = K \frac{\sqrt{E_p}}{n_1 n_2 E_g^3} F_2(\frac{\hbar \omega_1}{E_g}, \frac{\hbar \omega_2}{E_g}) \ ,
\label{eq1}
\end{equation}
where $K$ is a material independent constant, $E_p$ is the Kane energy parameter, $n_{1,2}$ are the refractive indices. And the function $F_2$ is given by
\begin{equation}
F_2(x_1, x_2) = \frac{(x_1+x_2-1)^{3/2}}{2^7 x_1 x_2^2} (\frac{1}{x_1} + \frac{1}{x_2})^2 \ ,
\label{eq2}
\end{equation}
which indicates that dramatically enhanced 2AP is available when either $\omega_1$ or $\omega_2$ becomes small. Note that for an indirect bandgap semiconductor, such as Si, the enhancement behavior has also been observed experimentally \cite{Poulvellarie2018PRAp, Zhang2015OE} and theoretically \cite{Cox2019SPIE}.

In our experiment, the pump wavelength was chosen to be 3070 nm for detecting the signal at 1550 nm, which correspond to photon energy of 0.4 eV and 0.8 eV. The combined photon energy of 1.2 eV exceeds the Si band gap of 1.12 eV, while the pump photon energy was lower the semiconductor midgap. In this case, the count rate recorded by a Si-APD could be expressed as
\begin{equation}
N_\text{total}  =  \beta P_1^2 + \gamma P_2^3 + \beta ' P_1 P_2  \ ,
\label{eq3}
\end{equation}
where $P_1$ and $P_2$ represent the signal and pump powers. In practical situation of sensitive detection, the faint signal is typically gated by a strong pump, $\textit{i.e.}$, $P_1 \ll P_2$. Hence the background noise induced by the pump is reduced to be
\begin{equation}
N_\text{D-3PA} = \gamma P_2^3  \ ,
\label{eq4}
\end{equation}
while the effective signal due the ND-2PA is 
\begin{equation}
N_\text{ND-2PA} = \beta ' P_1 P_2  \ .
\label{eq5}
\end{equation}
As a comparison to the detection scheme based on D-2PA of the signal, the enhancement factor $G$ for the resulting count rate can be defined as
\begin{equation}
G = N_\text{ND-2PA} / N_\text{D-2PA} = \beta ' P_1 P_2 / \beta P_1^2 \propto 1 / P_1 \ .
\label{eq6}
\end{equation}
Indeed, the strong pump field serves as a local oscillator for amplifying the photocurrent, which thus enables to detect extremely weak signals.
  
\section{Experimental setup}

Figure \ref{fig1} presents the experimental scheme. The initial light source was from a passively synchronized fiber laser system, which has been detailed in \cite{Zeng2019OL}. The involved Yb- and Er-doped fiber lasers (YDFL and EDFL) were mode-locked at the repetition rate of 15.8 MHz to deliver picosecond pulses at 1030 nm and 1550 nm, respectively. The power of generated pulses was then boosted by Yb- and Er-doped fiber amplifiers (YDFA and EDFA). The amplified dual-color pulses could facilitate the preparation of MIR pulses at 3070 nm by performing difference-frequency generation within a periodically-poled lithium niobate (PPLN) crystal. A delay line (Delay1) consisting of two fiber collimators was used to precisely tune the temporal overlap for reaching an optimized conversion efficiency. After a long-pass filter, the spectrally purified MIR light would serve as the pump source for implementing the subsequent 2PA-based detector. The other branch from the EDFL was used to prepare the near-infrared (NIR) signal source. The signal and pump powers could be continuously tuned by rotating the half-wave plate (HWP) before the polarization beam splitter (PBS) or the polarizer. Finally, the signal and pump sources were spatially combined by a dichroic mirror before being coupled into a ZrF4 single-mode patch cable (Thorlabs P3-23Z-FC-1). The fluoride fiber has a high intrinsic transmission above 97\% at the two relevant wavelengths. The mixed beam was steered into the a single-photon detector based on Si-APD (Excelitas SPCM-AQRH-54-FC). The count rate from the detector output was measured by a frequency counter (Tektronix, FCA3100). Note that another delay line (Delay2) was used to determinate the contributions of degenerate and non-degenerate responses \cite{Xu2019PTL}. More details about the experiment setup could be found in Appendix \hyperlink{appA}{A}.

\section{Results and discussion}

\begin{figure}[t!]
\centering
\includegraphics[width=0.9\columnwidth]{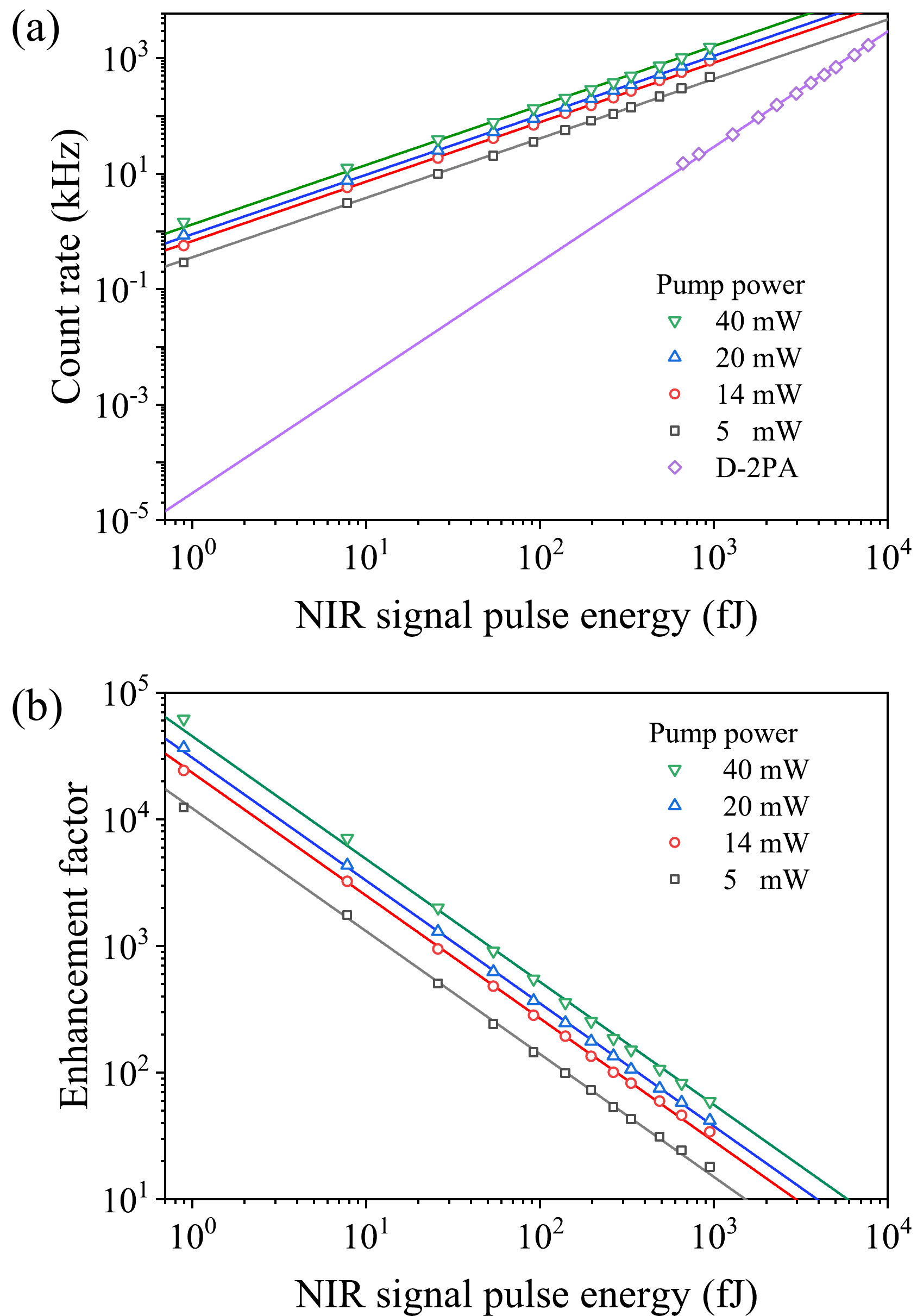}
\caption{Near-infrared detection performance based on the ND-2PA at the presence of different pump powers. (a) Count rate for the ND-2PA increases linearly with the input pulse energy. For comparison, D-2PA for the infrared signal is also included, which indicates a significant enhancement on the detection efficiency. (b) Enhancement factor for the recorded count rates due to ND-2PA effect with a comparison to the detection scheme based on D-2PA of signal alone.}
\label{fig3}
\end{figure}

Now we turn to characterize the photon-counting performance of the 2PA-based detector. Figure \ref{fig2}(a) presents the relevant transitions within the energy diagram of a silicon detector, which include D-2PA at 1550 nm, D-3PA at 3070 nm, and ND-2PA between them. The corresponding experimental data is shown in Fig. \ref{fig2}(b). The log-log plot shows a linear dependence between the recorded count rate and the input pulse energy. As expected, the fitted slopes for the cases of D-2PA, D-3PA, and ND-2PA were found to be 2, 3, and 1, respectively. Details about the fitting parameters were given in Appendix \hyperlink{appB}{B}. It can be seen that the pump-induced noise due to D-3PA is about six orders of magnitude lower than that due to D-2PA for a given pump power. This observation validates the efficacy of the MIR pumping approach to enhance the detection sensitivity. Practically, the pump power could be optimized to reach a background noise closed to the dark counts about 100 Hz defined by the Si detector itself. In the two aforementioned scenarios, the allowed power for MIR pumping is about 40 dB higher than that for NIR pumping, thus resulting in a much higher detection efficiency. In Fig. \ref{fig2}(b), an example of the linear response for the ND-2PA is given under a MIR pulse energy of 0.32 nJ. The detection floor about 1 kHz was defined by the pump-induced 3PA noise. As a result, the resolvable signal power could reach to the femtojoule level corresponding to about $8 \times 10^3$ photons per pulse.

Then we systematically investigated the responding behavior for the ND-2PA on the input signal power. In the experiment, the contribution of the ND-2PA was identified by the difference between the count rates corresponding to the full overlap and well separation for the signal and pump pulses. Figure \ref{fig3}(a) illustrates the expected linear dependence on the signal pulse energy for various settings of MIR pump power, as predicted by Eq. (\ref{eq5}). Also, the count rate would increase at the presence of a higher pump power, which implied a higher detection detection efficiency. Additionally, the ND-2PA count rate was greatly amplified due the presence of the pump in comparison to the D-2PA scheme. Following the definition by Eq. (\ref{eq6}), the enhancement factor would be inversely proportional to the signal power, as manifested in Fig. \ref{fig3}(b). Thanks to the significant suppression of the pump-induced noise, much lower signal power could be resolved. Consequently, an unprecedented enhancement factor about 6$\times 10^4$ was obtained for a signal pulse energy of 1 fJ under a 40-mW pump, thus showing a hundredfold improvement over previous demonstrations \cite{Knez2020LSA, Boitier2009APL, Xu2019PTL}, as listed in Appendix \hyperlink{appC}{C}.

\begin{figure}[t!]
\centering
\includegraphics[width=0.9\columnwidth]{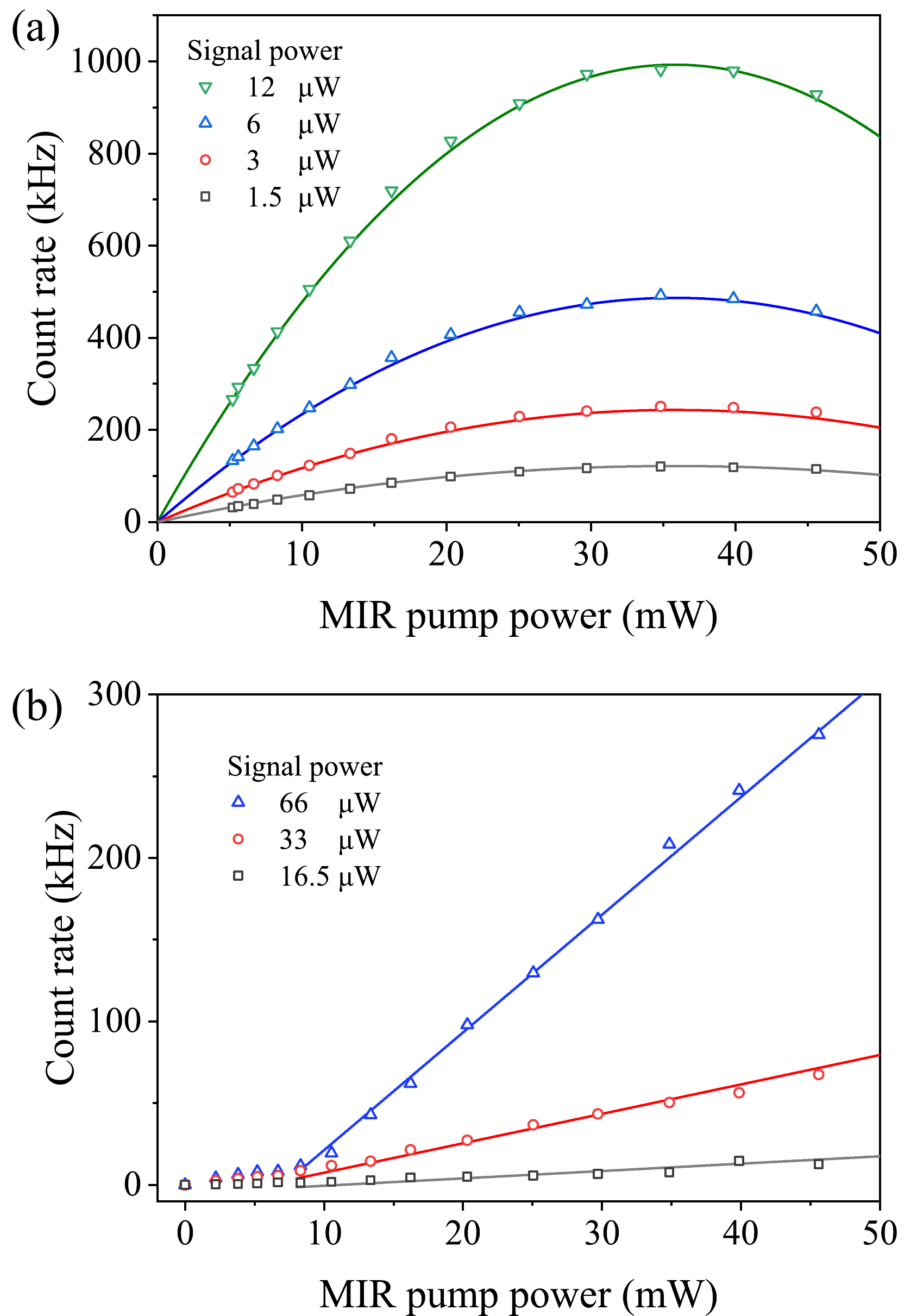}
\caption{(a) Count rates for the ND-2PA vary as the MIR pump power for various settings of signal power. The solid lines are given by the theoretical model in the text, where the dependence of the absorption coefficient on the pumping power has been taken into account. (b) Decrement of count rates for the D-2PA at 1550 nm with the increasing the MIR pump power. Note that the signal and MIR pulses are temporally separated for performing this measurement.}
\label{fig4}
\end{figure}

In the following, we proceed to investigate the evolution of ND-2PA count rate as a function of the MIR pump power. According to Eq. (\ref{eq5}), the count rate should be linearly proportional to the pump power. However, the expected linear dependence was only confirmed in the low-power pumping regime as shown in Fig. \ref{fig4}(a). Under higher pump powers, the count rate would tend to be saturated before starting to decrease dramatically. This phenomenon might be related to the thermal effect due the intensive MIR radiation. Indeed, we experimentally observed that the count rate would slowly decrease to a stabilized value when the pump was launched. Empirically, the ND-2PA coefficient $\beta ' $ could be expanded in a Taylor series as $\beta '=  \beta '^{(0)} + \beta '^{(1)} P_2$. The resulting expression of count rate due the ND-2PA is given by $N_\text{ND-2PA} = P_1(\beta '^{(0)} P_2+ \beta '^{(1)} P_2^2) $. Furthermore, identical behavior exhibited for the measured count rate due to D-TPA. In this case, the signal and pump pulses were temporally separated in order to exclude the nonlinear interaction between them. Similarly, the D-2PA coefficient $\beta$ could be expanded as $\beta =  \beta ^{(0)} + \beta ^{(1)} P_2$, which would result in a count rate $N_\text{D-2PA} =  P_1^2 (\beta ^{(0)} + \beta ^{(1)} P_2)$. The inhibiting effect due to the MIR pumping could be quantitatively characterized by $\Delta N_\text{D-2PA} = \beta ^{(1)} P_2 P_1^2$. As given in Fig. \ref{fig4}(b), the decrement of the D-2PA count rate for the signal was indeed more pronounced as augmenting the pump power. Further studies showed that this inhibiting phenomenon could also be observed for single-photon absorption at 1030 nm and three-photon absorption at 3070 nm. Therefore, the exhibiting suppression of the count rate might be ascribed to the variation of the reverse-biased voltage for the optimized operation of the single-photon detector.

\begin{figure}[b!]
\centering
\includegraphics[width=0.95\columnwidth]{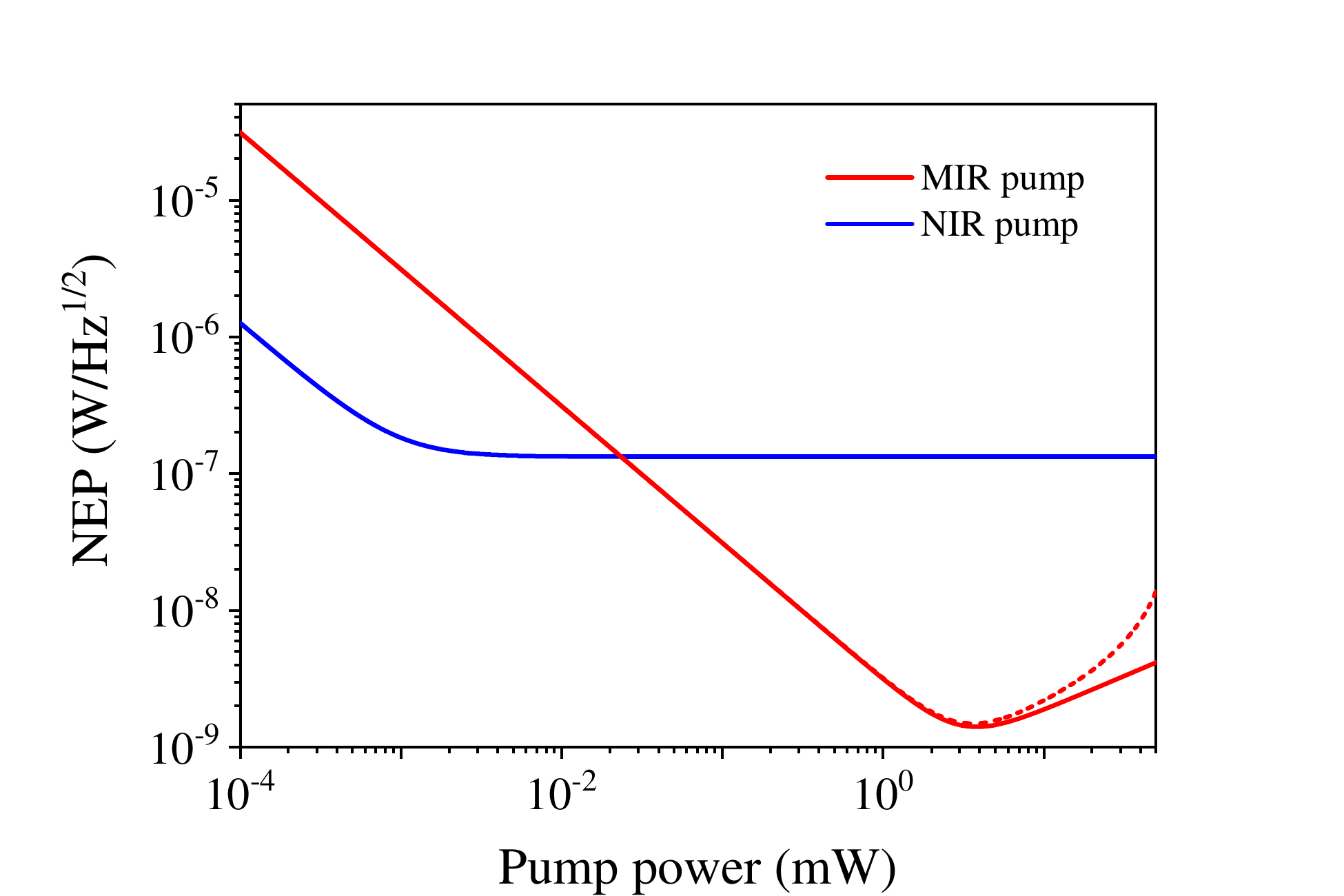}
\caption{Noise equivalent power (NEP) as a function of the pump strength in the cases of using two typical pump wavelengths at 3070 nm and 1550 nm. Comparatively, much lower NEP could be obtained in the MIR pumping scenario. Note that the dashed line represents the model by including the thermal effect due the strong pump intensity.}
\label{fig5}
\end{figure}

Finally, we have analyzed the noise equivalent power (NEP), which is an important figure of merit for characterizing the sensitivity of optical detectors. It is desirable to have an NEP as low as possible, which corresponds to a lower noise floor and hence a more sensitive detector. Particularly, the NEP for a photon counter is typically defined as \cite{Kang2020PTL}: NEP = $h \nu \sqrt{2 N_\text{b}}/\eta$, where $h \nu$ represents the signal photon energy, $N_\text{b}$ denotes the background count rate, and $\eta$ is the detection efficiency. For a 2PA-based detector, $\eta$ is given by the ratio between the detected count rate $N_\text{ND-2PA}$ and input photon flux $N_1$: $\eta = \frac{N_\text{ND-2PA}} {N_1}  = \frac{\beta ' P_1 P_2}{P_1/ h \nu} = \beta ' P_2 h \nu$. And $N_\text{b}$ is comprised of the dark noise $N_0$ of the Si-APD and the pump-induced noise: $N_\text{b}^\text{MIR} =  N_0 + \gamma P_2^3$ and $N_\text{b}^\text{NIR} =  N_0 + \beta P_2^2$, which correspond to MIR and NIR pumping, respectively. Therefore, the expressions of NEP for the two cases are given by
\begin{equation}
\begin{split}
\text{NEP}^\text{MIR}& =  \frac{\sqrt{2(N_0 + \gamma P_2^3)}}{\beta ' P_2} \ , \\
\text{NEP}^\text{NIR}& =  \frac{\sqrt{2(N_0 + \beta P_2^2)}}{\beta ' P_2}  \ .
\label{eq8}
\end{split}
\end{equation}
At the weak-pumping condition of $P_2 \to 0$, the background noise is dominated by the detector dark noise, which leads to NEP $\propto P_2^{-1}$. At the other extreme condition with $P_2 \to + \infty$, $\text{NEP}^\text{MIR}  \propto P_2^{1/2}$ and $\text{NEP}^\text{NIR}$ approaches to be constant. These evolutions were illustrated in Fig. \ref{fig5}, which indicates a significant improvement of NEP under MIR pumping by two orders of magnitude. Note that the wavelength in the NIR-pumping scenario was set to 1550 nm. Actually, other wavelengths below 2200 nm would be expected to exhibit similar behavior, albeit with tolerable deviations. The parameters used in the simulation were obtained from the fitted values for the experimental data shown in Figs. \ref{fig3} and \ref{fig4}.

It is worth noting that the MIR-pumping strategy should in principle be applicable for broadband sensitive detection at infrared. For instance, wavelengths up to 1.7 $\mu$m is applicable for the operational MIR pump at 3 $\mu$m. Furthermore, a spectrum over 1.2 - 2 $\mu$m could be covered with an adapted pump wavelength about 2.4 $\mu$m. This particular wavelength can be readily accessed by MIR lasers based on Cr$^{2+}$:ZnSe or Cr$^{2+}$:ZnS \cite{Ma2019APR}.

\section{Conclusion}

In summary, we have demonstrated sensitive infrared detection in the Si-APD using ND-2PA based on MIR pumping scheme. In this non-degenerated regime, the improvement of nonlinear absorption coefficient for the signal and the elimination of second-harmonic nosies of the pump enabled us to increase the detection efficiency and reduce the background noise. Consequently, femtojoule-level pulse energy could be identified, which represented a record sensitivity for 2PA-based detectors. The corresponding enhancement factor of the ND-2PA count rate unprecedentedly reached to about $10^5$ comparing to the scheme based on D-2PA.

To go beyond the achieved sensitivity, lower pump photon energy could be employed with a MIR wavelength over 3300 nm. The 3PA noise might thus be removed, which holds potential to ultimately realize single-photon sensitivity as theoretically predicted \cite{Hayat2008PRB}. Furthermore, the detection performance would also be enhanced by using waveguide geometry or micro-cavity structure with the help of the tight spatial confinement and long nonlinear interaction length \cite{Poulvellarie2018PRAp, Zhang2015OE}. It is worthy noting that the efficacy of our implemented configuration could immediately facilitate sensitive infrared imaging via a Si electron-multiplying CCD, thus favoring desirable features of high spatial resolution and high-speed frame rate. Additionally, the gated detection by pulsed pumping provides a high-precision temporal resolution for sensitive infrared detection and imaging, which may be useful for relevant applications, such as time-resolved molecular spectroscopy and dynamic photoluminescence analysis.

\section*{Acknowledgments}
This work was supported by National Key Research and Development Program (2018YFB0407100), Science and Technology Innovation Program of Basic Science Foundation of Shanghai (18JC1412000), Program for Professor of Special Appointment (Eastern Scholar) at Shanghai Institutions of Higher Learning, National Natural Science Foundation of China (11621404, 11727812), Shanghai Municipal Science and Technology Major Project (2019SHZDZX01).

\section*{APPENDIX A: Experimental setup}
\label{appendixA}
\hypertarget{appA}{The} whole experimental setup mainly consisted of three parts, corresponding to realize synchronized fiber lasers, prepare signal and pump sources, and implement 2PA-based detection. The synchronized  laser system was based on the Er-doped and Yb-doped fiber lasers in a master-slave layout. The master pulses at 1550 nm were injected into the laser cavity of the YDFL based on nonlinear amplifying loop mirror (NALM). The induce a periodic nonreciprocal phase shift via the cross-phase modulation effect within the NALM initiated the synchronous mode-locking of the slave laser, which resulted in a passive locking of the relative repetition rates between the two lasers \cite{Zeng2019OL}. The pulse durations of the YDFL and EDFL were measured to be 5 ps and 36 ps by using an auto-correlator. The relative timing jitter was measured to be about 20 fs, which was negligible comparing to the pulse duration. In order to prepare the mid-infrared pump source, we performed difference-frequency generation in a periodically-poled lithium niobate (PPLN) crystal. The pulse duration of the generated MIR field was inferred to be 5 ps based on the cross-correlation technique. A long-pass filter with a cut-off wavelength of 2.4 $\mu$m was used to removed the near-infrared fields. The MIR power could be varied by using a combination of a nanoparticle linear film polarizer (Thorlabs, LPNIRA050) and low-order half-wave plate (Thorlabs, WPLH05M-2940), which would serve as the pump source for the subsequent 2PA-detection. In parallel, the signal source was from the attenuated light at 1550 nm. The near-infrared signal was filtered by a band-pass filter centered at 1550 nm with a bandwidth of 40 nm. A delay line was inserted into signal path for temporally tune the overlap between the dual-color pulses. Then, the signal and pump sources were spatially combined by a dichroic mirror before being coupling into a fluoride single-mode fiber. The coupling efficiencies for both beams were about 45\%.

\section*{APPENDIX B: Fitted parameters}
\label{appendixB}

\begin{figure*}[t!]
\centering
\includegraphics[width=0.9\textwidth]{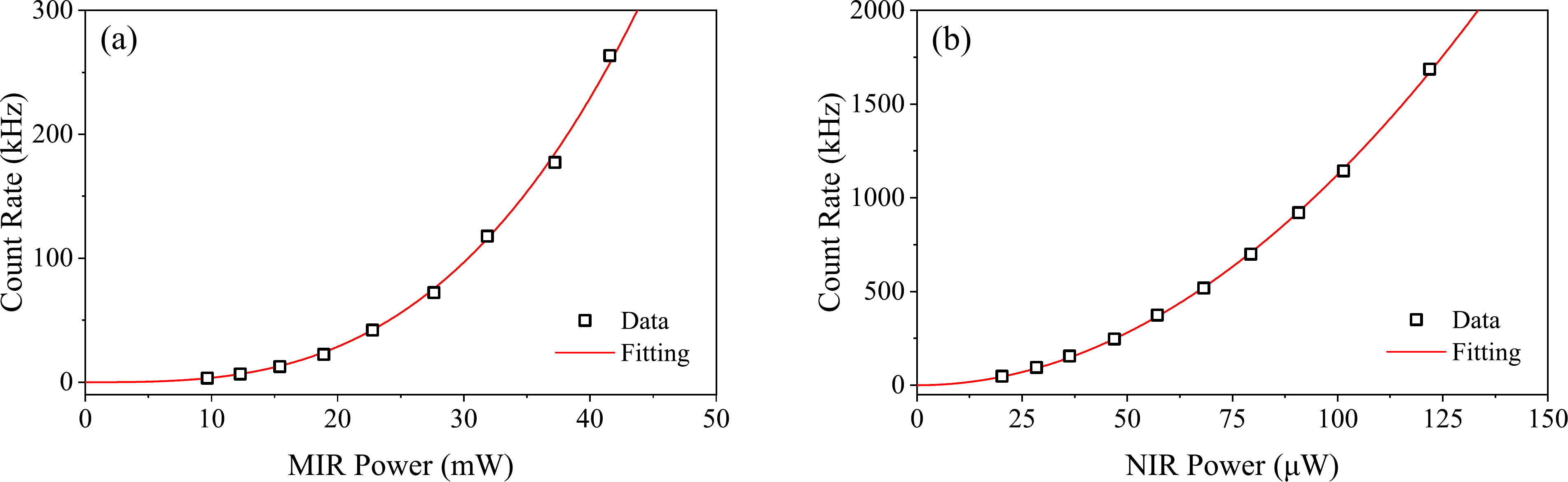}
\caption{Count rates as a function of the used power for the MIR-pumping (a) and NIR-pumping (b) scenarios. The fitted lines are used to estimate the values of $\beta$ and $\gamma$ for D-2PA and D-3PA processes in the Si-APD.}
\label{fig6}
\end{figure*}

\hypertarget{appB}{The} parameters for modeling the experimental data was summarized in table \ref{table1}. Specifically, the D-2PA and D-3PA coefficients were obtained from the quadratic and cubit fittings as shown in Fig. \ref{fig6}. 

\begin{table}[h!]
\centering
\caption{Fitted parameter for the involved 2PA and 3PA processes within a Si-APD.}
\centering
\label{table1}
\begin{tabular}{p{0.18\linewidth}p{0.18\linewidth}p{0.18\linewidth}p{0.18\linewidth}p{0.14\linewidth}}
\hline
$\beta^{(0)}$ [Hz/W$^2$] & $\beta^{(1)}$ [Hz/W$^3$] & $\beta '^{(0)}$ [Hz/W$^2$] & $\beta'^{(1)} $ [Hz/W$^3$] & $\gamma$ [Hz/W$^3$] \\
\hline
1.12$\times10^{14}$  & -1.65$\times10^{15}$ & 4.53$\times10^{12}$ & -6.33$\times10^{13}$ & 3.59$\times10^{9}$ \\
\hline
\end{tabular}
\end{table}

Based on the these fitting parameters, the peak detection efficiency of the implemented detector based on ND-2PA is given by $\eta_\text{ND-2PA} = \beta ' P_2 h \nu = 1.04 \times 10^{-8}$ at the presence of 35-mW pump power. However, the minimum NEP is optimized at the pump power of 4 mW, leading to $\text{NEP}^\text{MIR} =  \frac{\sqrt{2(N_0 + \gamma P_2^3)}}{\beta ' P_2} = 1.4 \times 10^{-9} \ \text{W/Hz}^{1/2}$. 

\section*{APPENDIX C: Performances for comparison}
\label{appendixC}

\hypertarget{appC}{As} shown in table \ref{table2}, we have listed related works on 2PA-based detection for direct comparison. It can be seen that the ND-2PA based on MIR pumping favors in enhancing the performance on detection sensitivity.

\begin{table*}[t!]
\centering
\caption{Comparison of passive synchronization performance for mode-locked fiber lasers with comparable pulse durations. $\lambda_s$ and $\lambda_s$ indicate the employed signal and pump wavelengths. The photon energy of the pump and the band gap of the detector are denoted as $E_p$ and $E_g$, respectively. $E_\text{min}$ represents the minimum detectable energy. $G_\text{max}$ is the peak enhancement factor for the ND-2PA comparing to the detection based on D-2PA.}
\centering
\label{table2}
\begin{threeparttable}
\begin{tabular}{p{0.1\linewidth}p{0.1\linewidth}p{0.1\linewidth}p{0.1\linewidth}p{0.1\linewidth}p{0.1\linewidth}p{0.1\linewidth}p{0.1\linewidth}}
\hline
Ref. & $\lambda_s$ [$\mu$m] & $\lambda_p$ [$\mu$m] & $E_p$ [eV] & Detector & $E_g$ [eV] & $E_\text{min}$ [fJ] & $G_\text{max}$ \\
\hline
This work & 1.55 & 3.07 & 0.4 & Si & 1.12 & $1$ & 6$\times 10^{4}$ \\
\cite{Fishman2011NP} & 5.6 & 0.39 & 3.19 & GaN & 3.28 & $10^{6}$ $^\text{[a]}$ & $10^{3}$ \\
\cite{Knez2020LSA} & 3.39 & 1.48 & 0.84 & Si & 1.12 & 200 & $10$ \\
\cite{Boitier2009APL} & 1.55 & 1.9 & 0.65 $^\text{[b]}$ & GaAs & 1.42 & $10^{7}$ $^\text{[c]}$ & $300$ \\
\cite{Xu2019PTL} & 1.85 & 1.55 & 0.8 & Si & 1.12 & 16 & $20$ \\
\hline
\end{tabular}
\begin{tablenotes}
\item[a, c] These experiments were operated with a continuous-wave signal input.\\
\item[b] Although the pump photon energy is below the midgap of GaAs detector, the 2PA noise due to pump was dominated in the experiment.
\end{tablenotes}
\end{threeparttable}
\end{table*}

\end{document}